# Statistical uncertainty analysis for small-sample, high log-variance data: Cautions for bootstrapping and Bayesian bootstrapping

Barmak Mostofian and Daniel M. Zuckerman[*]

Department of Biomedical Engineering, School of Medicine,

Oregon Health & Science University, Portland, Oregon

[*]zuckermd@ohsu.edu

**Abstract**

Recent advances in molecular simulations allow the evaluation of previously unattainable observables, such as rate constants for protein folding. However, these calculations are usually computationally expensive and even significant computing resources may result in a small number of independent estimates spread over many orders of magnitude. Such small-sample, high "log-variance" data are not readily amenable to analysis using the standard uncertainty (i.e., "standard error of the mean") because unphysical negative limits of confidence intervals result. Bootstrapping, a natural alternative guaranteed to yield a confidence interval within the minimum and maximum values, also exhibits a striking systematic bias of the lower confidence limit in log space. As we show, bootstrapping artifactually assigns high probability to improbably low mean values. A second alternative, the Bayesian bootstrap strategy, does not suffer from the same deficit and is more logically consistent with the type of confidence interval desired. The Bayesian bootstrap provides uncertainty intervals that are more reliable than those from the standard bootstrap method, but must be used with caution nevertheless. Neither standard nor Bayesian bootstrapping can overcome the intrinsic challenge of under-estimating the mean from small-size, high log-variance samples. Our conclusions are based on extensive analysis of model distributions and re-analysis of multiple independent atomistic simulations. Although we only analyze rate constants, similar considerations will apply to related calculations, potentially including highly non-linear averages like the Jarzynski relation.

**Introduction**

In some sense, the *error bars* computed for observables of interest can be seen as the primary goal of quantitative simulation. After all, an unreliable mean value can be useless or even misleading. In this report, we describe challenges and partial solutions for generating reliable confidence intervals based on small-sample, high *log*-variance data sets. High log-variance data are characterized by values spanning multiple orders of magnitude. Our motivation stems from the domain of molecular simulation, but our analysis is intended to be very general.

In the study of large biomolecules such as proteins, all-atom molecular dynamics (MD) simulations remain extremely expensive because protein systems exhibit functional behaviors spanning timescales from μs to sec and beyond,[1-2] whereas most MD studies are limited to the μs scale.[3] Although numerous sophisticated sampling algorithms have been developed,[4-7] these generally still require significant computing resources,[8-10] with the ultimate result that computational predictions tend to be characterized by significant variance.

One strategy for the direct simulation of long-timescale events is the weighted ensemble (WE) path sampling method[11-14] that has recently been employed for steady-state simulations of large conformational changes[15-16] and for ligand (un)binding.[17] The WE approach promotes the unbiased simulation of less probable states using a strategy of pruning and replicating trajectories of (dis)interest, and allows estimation of the rate constant for the simulated process.[18-19] However, due to the expense of running a large number of trajectories, only a small number of independent WE simulations can be performed. The result is a set of rate constants that may be spread over multiple orders of magnitude, i.e., they have a high log-variance, or variance of base-10 logarithms significantly exceeding one. The mean value of such data sets will be dominated by only a few values, and we would like to be able to assess the quality of such estimates *via the reliability of their associated error bars*.

In the broader context of rate estimation for biomolecular processes, we would like to estimate rate constants within one order of magnitude, given today's computing capabilities. A larger uncertainty would compromise the ability to discriminate between fast and slow processes, which arguably is the key goal of rate estimation. On the other hand, realizing lower uncertainty for complex systems not only is extraordinarily challenging but also may not be consistent with modern fixed-charge molecular force fields (the governing models of molecular simulation) which can exhibit inaccuracies of 1 kcal/mol or more for free-energy calculations dominated by energy minima.[20-21] Force fields are not parametrized to kinetic observables which will be significantly influenced by energy barriers.

Error bars should represent confidence intervals, and we would like to assess the reliability of confidence intervals obtained for small-sample, high log-variance data sets. An X% confidence interval is defined as a range that covers the mean of the underlying distribution X% of the time based on multiple repeated measurements.[22] Of course, to obtain a reliable estimate of any statistic for an underlying distribution, the empirical data must be representative of that distribution. If the empirical data set is small, derived statistics can be expected to suffer from large uncertainties. The confidence interval itself is such a derived statistic which, as noted, is critical for assessing simulation results.

Here we attempt to quantify the reliability of confidence intervals for the mean value derived from small data sets with high log-variance. Common methodologies can easily fall short. The standard error of the mean of a set of independently obtained data values is the most frequently employed measure to assess the accuracy of the mean and it can be used to define a certain confidence interval for the mean;[23] for example, two standard errors of the mean roughly correspond to a 95% confidence interval.[24] However, for small-sample, high log-variance sets of rate-constant estimates, which are positive definite, the standard error often leads to lower interval boundaries that are negative and thus unphysical. We therefore do not use the standard-error approach here.

Bootstrap analysis is also very common and yields confidence intervals which are guaranteed not to fall outside the minimum and maximum value. The bootstrap approximates an underlying distribution by resampling data in a nonparametric way – i.e., without assumptions regarding the functional form of the distribution but assuming the given sample is representative for its sample size.[25] With the repeated resampling of the empirical data, each yielding a plausible sample mean, a bootstrap distribution of such mean values is generated that can be used to estimate the precision of the mean based on the original empirical data set. The procedure constructs confidence intervals which implicitly address the question, *"If the initial sample is representative, what range of statistics (e.g., sample means) does one expect to observe?"* The bootstrap has become a widely accepted approach for providing confidence intervals but some limitations have been noted.[26-27]

The Bayesian bootstrap is much less common in molecular simulation analyses and uses a different strategy to assess the uncertainty of a statistic's estimate.[28] Unlike the standard (frequentist) bootstrap method, the Bayesian approach generates random "posterior" samples of non-uniform *probabilities* for a multinomial distribution of the empirical data – i.e., alternative distributions, each yielding its own mean. A "credibility region" – the Bayesian analog of the confidence interval – can be derived from the posterior distribution of means by bracketing the values in analogy to ordinary bootstrapping (e.g., a 95% credibility region). Although it is a resampling method, the Bayesian bootstrap is different from the standard bootstrap because it does not resample the empirical data but instead samples multiple alternative posterior distributions – each yielding its own mean. In simple terms, the Bayesian credibility region addresses the question, *"Based on the sample obtained, what range is expected for the true mean of the underlying distribution?"* Note the stark difference of this question from that for the standard bootstrap.

We assess the usefulness of the standard and Bayesian bootstrap analyses for a variety of different distributions that may be spread over several orders of magnitude, such as the power law or exponential distributions. In particular, we generate many synthetic data sets along with a confidence interval *and* credibility region for each (see Figure 1) and examine whether and how these bracket the "true" distribution mean. This allows us to systematically assess the differences between the standard and the Bayesian bootstrap depending on the underlying distribution, the size of the data samples, and the range of the confidence interval or credibility region. Subsequently, we perform the same analysis on protein folding rate constants originally obtained from multiple independent WE simulations.[16] Importantly, we use these simulation data sets as representative examples of "ground truth" from which to generate synthetic data to quantify the reliability of the predicted intervals/regions.

The results show that the credibility regions from the Bayesian bootstrap tend to bracket the true mean much more tightly, i.e., by comparison, the confidence interval from the standard bootstrap is orders of magnitude too large (in log space) in many cases. Hence, the standard bootstrap inaccurately describes the level of precision within the data on the desired log scale (such as for rate constants). We further show that the log variance, the standard deviation of the logarithms of data values, is a simple but reliable indicator for when the Bayesian bootstrap is expected to perform better.

The reason for the imprecise confidence intervals of small-sample, high log-variance data is because significant fractions of standard bootstrap samples lack enough large values. This can be understood from a simple example. Imagine a small set of data that only consists of zeros and ones: for instance, four evaluations resulted in the value 0 and one result assumes the value of 1. In fact, this small set of binary results is a good model for the rate constants from WE simulations, for which, as noted before, the mean value is dominated by a small subset of the results. During the standard bootstrap of this small data set, however, there would be many (re)samples lacking the value of 1, leading to bootstrap sample means of 0, which, in turn, yield a confidence interval that is biased toward very small values. The Bayesian bootstrap strategy remedies this strong bias.

Nevertheless, neither bootstrap method, we emphasize, can overcome the intrinsic limitations of very small sample sizes. More specifically, for the WE rate constants, the nominal 95% intervals/regions of both methods under-estimate the "true" rate constant mean in ~30% of cases, and up to even ~55% for data sets of comparable size from continuous distributions.

**Methods**

*Standard Bootstrap.* The standard bootstrap is a method for statistical inference based on random resampling, i.e. drawing with replacement, of empirical data.[25] Each sample can be evaluated with respect to a certain statistic, e.g. the sample mean. The repeated generation of such random samples allows quantification of the reliability of the estimate for that statistic in a frequentist way and without any prior assumptions regarding the functional form of the true distribution. Let $\boldsymbol{x} = (x_1, x_2, \ldots, x_n)$ be the empirical data set and $\bar{x} = \frac{1}{n}\sum_{i=1}^{n} x_i$ its mean. The bootstrap algorithm randomly resamples from $\boldsymbol{x}$ to generate $\boldsymbol{x}^* = (x_1^*, x_2^*, \ldots, x_n^*)$. This operation is repeated $B$ times, producing a set of bootstrap samples or replicates $(\boldsymbol{x}_1^*, \boldsymbol{x}_2^*, \ldots, \boldsymbol{x}_B^*)$ from which a set of bootstrap sample means $\bar{\boldsymbol{x}}^* = (\bar{x}_1^*, \bar{x}_2^*, \ldots, \bar{x}_B^*)$ can be derived. Moreover, bootstrap percentiles can be derived to define a suitable confidence interval, as follows.[25] Let $CDF_{\bar{x}^*}$ be the cumulative distribution function of all $\bar{x}_i^*$; then a 95% confidence interval is obtained by $[\bar{x}_{lo}, \bar{x}_{up}] = [CDF_{\bar{x}^*}^{-1}(0.025), CDF_{\bar{x}^*}^{-1}(0.975)]$, which is supposed to cover $\bar{x}$, and therefore $\mu$, 95% of the time.

*Bayesian Bootstrap.* The Bayesian bootstrap[28] is a Bayesian analogue of the bootstrap method. It differs from the standard method in that, at any bootstrap iteration, a multinomial *distribution* is resampled but the given empirical data are always the same. Specifically, each Bayesian bootstrap replication represents a different multinomial distribution in which each of the empirical $x_i$ values is assumed to have been chosen with probability $\pi_i^*$ based on a set of random *parameters* $\boldsymbol{\pi}^* = (\pi_1^*, \pi_2^*, \ldots, \pi_n^*)$, normalized according to $\sum_{i=1}^{n} \pi_i^* = 1$. The multinomial distribution of the empirical $x_i$ values is a natural choice for Bayesian bootstrapping given a fixed set of discrete data from an unknown distribution: it uses all the empirical data values, and only these. Each round of bootstrap produces a sample mean $\bar{x}^* = \sum_{i=1}^{n} \pi_i^* x_i$. If we let $CDF_{\bar{x}^*}$ be the cumulative distribution function of all $\bar{x}_i^*$, then a 95% credibility region is obtained by $[\bar{x}_{lo}, \bar{x}_{up}] = [CDF_{\bar{x}^*}^{-1}(0.025), CDF_{\bar{x}^*}^{-1}(0.975)]$, which is supposed to cover $\mu$, 95% of the time. Note that the procedure just outlined implicitly uses a prior $P(\{\pi_i^*\}) = 1/\prod_{i=1}^{n} \pi_i^*$ that is the inverse of the multinomial likelihood.[28]

To resample multinomial distributions, we generated sets of *n* normalized random probabilities $\pi_i^*$ using the method described by Rubin.[28] The *n* values are derived as the differences (or "gaps") between succeeding pairs in an ordered list of *n+1* real numbers between 0 and 1, generated by combining the values 0 and 1 with *n-1* uniform variates. This procedure leads to the correct distribution of $\pi_i^*$ values (Figure S1A), in contrast to a naïve approach of simply normalizing *n* uniform variates (Figure S1B).

The code for this Bayesian bootstrap procedure can be obtained from *https://github.com/ZuckermanLab/BayesianBootstrap*.

*Data Sources and Data Generation.* We first applied the standard and the Bayesian bootstrap to data randomly sampled from different common continuous distributions (see Table 1) in order to reach general conclusions on their performance. Beyond that, we continued with rate constants that are the actual results of previous WE simulations (see Table 2).[16]

We employed several different types of continuous distributions with variable widths (listed in Table 1): data uniformly spread over several orders of magnitude between 0 and 1, data distributed according to the power law $x^{-\alpha}$ with $0 < x \leq 1$ or with $x > 0$, data exponentially distributed according to $exp(-\lambda x)$, and data normally distributed according to $exp\frac{-(x-\mu)^2}{2\sigma^2}$.

To assess the performance of the standard and the Bayesian bootstrap methods on each distribution, we generated a large number ($N = 1000$) of synthetic data sets by repeated sampling from each distribution. The size of these synthetic data samples ranged from $n = 5$ to $n = 1000$. We performed both the standard and the Bayesian bootstrap on each synthetic data set leading to a confidence interval and credibility region in each instance as displayed in Figure 1. The number of bootstrap resampling or replications was $B = 10,000$ in both procedures. Note $B \neq N$ and that $B$ bootstrap repeats were performed on *each* of the $N$ synthetic data sets. The performance of the two different bootstrap methods for error analysis was evaluated by assessing the ability of the corresponding nominal 50%, 65%, 80%, or 95% confidence interval and credibility region to cover the true mean, $\mu$. Ratios of under- or over-estimation were obtained by counting how frequently (out of $N = 1000$) the upper limit was smaller or the lower limit was larger than the true distribution mean. Moreover, the degree of log variance of the data was described by the standard deviation of the base-10 logarithm of synthetic data set means, $\sigma_{\log_{10}(\bar{x})}$.

The simulation data were obtained from WE folding simulations of two different proteins using the WESTPA software package.[29] These two proteins were the N-terminal domain of the ribosomal protein L9 (NTL9, PDB: 2HBB), referred to as System A, and the immunoglobulin-binding domain of Protein G (PDB: 1PGA), or System B. With the Hill relation,[30] WE allows for the calculation of the folding rate constant from the probability flux entering the folded state of a protein. Here, we use a subset of the folding rate constants ($n = 15$ for System A and $n = 13$ for System B, listed in Table 2), which were obtained from the final simulation time of independent WE simulations,[16] as the two underlying (ground truth) data sets, with true mean $\mu$, for the bootstrap error analysis. The aggregate WE simulation times accounting for all trajectories were ~60 μs (System A) and ~200 μs (System B), respectively.

We derived $N$ synthetic data sets ($N = 1000$) by sampling with replacement from the two example WE data sets, which were assumed to be the ground truth (i.e., the correct distribution of observables, implying the WE sample mean is the true mean). Each synthetic data set was of the same size as the corresponding true data set ($n = 15$ for System A and of $n = 13$ for System B, respectively). We assessed the performance of both bootstrap methods for nominal confidence/credibility coverages of 95%.

## Results & Discussion

**Data sampled from continuous distributions**

Our analyses compare the standard and Bayesian bootstrap error estimation for small-sample, high log-variance data. Hence, we apply these methods to synthetic data sets randomly generated from several ("true") underlying continuous distributions spanning several orders of magnitude, as listed in Table 1. These distributions are (i) a log-scale uniform distribution over $k$ orders of magnitude between 0 and 1, i.e. $\log_{10}(x) \sim k \times \mathcal{U}(-1,0)$, where $\mathcal{U}(-1,0)$ is the uniform distribution between -1 and 0, (ii) power law distributions, i.e. $x \sim x^{-\alpha}$, either with $0 < x \leq 1$ or with $x > 0$, (iii) an exponential distribution, i.e. $x \sim \exp(-\lambda x)$, and (iv) a normal distribution, i.e. $x \sim exp\frac{-(x-\mu)^2}{2\sigma^2}$. The parameters $k, \alpha, \lambda$, and $\sigma$ were chosen such that the underlying distributions in this study, and thus the synthetic data sampled, have different log-variances so that we can assess their effect on the bootstrap performance.

The 95% credibility regions from the Bayesian bootstrap (red bars) consistently cover a smaller range, without significant loss of accuracy, compared to the 95% confidence intervals from the standard bootstrap (blue bars) as shown in Figure 2. For clarity, only 50 out of $N = 1000$ synthetic data samples are exhibited for three distributions, which are expected to have a relatively large log-variance, i.e., (A) the log-scale uniform distribution with $k = 20$, (B) the power law distribution with $\alpha = 0.9$, and (C) the exponential distribution with $\lambda = 1$ (the normal distributions, which are not expected to have large log-variance, are discussed in the SI). Note that, in these examples, the bootstrap performance is examined at a high confidence/credibility level (95%) for a fairly small size of synthetic data ($n = 10$) from distributions with relatively high log-variances (further examples with smaller values for $k, \alpha$, and $\lambda$ are shown in Figure S2).

The data sets sampled from the log-scale uniform distribution have mean values that vary by several orders of magnitude, whereas those from the power law distribution vary by about two orders, and those from the exponential distribution by less than one order of magnitude. This order-of-magnitude variance is captured by the standard deviation of the logarithm of $\bar{x}$, $\sigma_{\log_{10}(\bar{x})}$, suggesting that a value of $\sigma_{\log_{10}(\bar{x})} = 0.2 - 0.5$ is sufficient for them to differ from each other by more than one order of magnitude.

The lower limits of the standard bootstrap confidence intervals can reach very small values (e.g., $\bar{x}_{lo}$ < 1e-09 for the log-scale uniform distribution), which is clearly a bias since the corresponding mean value is many orders of magnitude larger, yielding error bars that are several orders of magnitude wide (~8 orders for the log-scale uniform distribution). In contrast, the Bayesian bootstrap credibility regions do not have this strong bias toward very small data values and thus provide a more reliable measure for the uncertainty of the sample mean value, $\bar{x}$.

To quantify the performance of the uncertainty ranges, we first analyzed how frequently their upper limits ($\bar{x}_{up}$) were smaller than the true mean, which we refer to as under-estimation, and how frequently their lower limits ($\bar{x}_{lo}$) were larger than the true mean, referred to as over-estimation. Averaged over all synthetic data sets ($N = 1,000$), both bootstrap approaches under-estimate the true mean by ~55% for the log-scale uniform distribution, by ~23% for the power law distribution, and by ~12% for the exponential distribution. However, they over-estimate the true mean of all distributions by less than 3%. The expected values for both under- and over-estimation of a 95% interval or region are actually 2.5%. Based on these examples, the nominal 95% coverage range effectively only covers the three distribution types at ~42% (log-scale uniform), ~75% (power law), and ~85% (exponential), respectively.

Interestingly, reducing the log-scale uniform spread of data from $k = 20$ to $k = 5$ orders of magnitude or the power law exponent from $\alpha = 0.9$ to $\alpha = 0.1$, significantly improves the under-estimations to ~25% (Figure S2A) or only ~5% (Figure S2B), respectively, which is in line with the smaller variance of the corresponding sample means in both distributions. However, reducing $\lambda$ from 1 to $10^{-6}$, does not have any measurable effect on the actual bootstrap coverage of the exponential distribution (Figure S2C), consistent with a constant value of $\sigma_{\log_{10}(\bar{x})}$, which is a consequence of the fact that the scale parameter of any exponential distribution, $\lambda^{-1}$, equals its mean and standard deviation. These results show that both the standard and the Bayesian bootstrap are less accurate in assigning uncertainty to mean values derived from continuous high log-variance distributions over several orders of magnitude.

To further investigate the effect of the size $n$ of synthetic data sets on the bootstrap under- and over-estimations of the true mean $\mu$, we systematically examined uncertainty ranges at several nominal coverages (50%, 65%, 80%, and 95%) with increasing $n$. Figures 3 and S3 illustrate the effect of the size of synthetic data sets: the percentage of under-estimation, which, for small samples, is clearly beyond the corresponding expected value (indicated by horizontal dashed lines at 25%, 17.5%, 10%, and 2.5% in Figures 3 and S3), converges to the expected value with increasing $n$. This is particularly true for the log-scale uniform distributions, which have a relatively high log-variance (Figures 3A and S3A). This inaccuracy for small-sample, high log-variance data is an intrinsic limitation of the bootstrap as it occurs at every tested nominal coverage of the true mean value, thus leading to a smaller effective coverage in the error analysis.

Interestingly, the degree of over-estimation for small $n$ seems to be *below* its expected value (same values as for the expected under-estimations mentioned above), particularly for high log-variance distributions at low nominal coverage. This occurs mainly because the uncertainty is biased toward smaller values in those simulations, as discussed before, thus reducing the effective number of over-estimations.

Figures 4 and S4 further quantify the methods' performance using the CDF of the upper and lower limits measured as a ratio to the true distribution mean for different relatively small sizes of synthetic data sets ($n$ = 5, 10, 25, 50) at several nominal coverages. The bias of the confidence intervals from the standard bootstrap toward small values is visualized by the strong increase of the corresponding CDFs for $\log(\bar{x}_{lo}/\mu) < 0$. The degree of under-estimation of both methods is shown by the increase of the CDFs for $\log(\bar{x}_{up}/\mu) < 0$. The changing shape of the sigmoidal curves with changing parameters is in line with the preceding results on bootstrap bias and effective coverage. For instance, the decrease in under-estimation of $\mu$ with increasing $n$ and with a lower distribution log-variance can be reconciled by comparing the relatively high CDF (~72%) at $\log(\bar{x}_{up}/\mu) = 0$ for $n$ = 5 from the log-scale uniform distribution (Figure 4A, upper panel, left column) to the corresponding CDF value for larger $n$ and/or for samples from the other distributions (Figures 4B, C). In particular, the opening gap between the standard and the Bayesian bootstrap CDFs for $\log(\bar{x}_{lo}/\mu) < 0$ at $n$ = 5 to 50 in Figure 4A (lower panel), and to some extent in Figure S4A (lower panel), illustrates the greater suitability of the Bayesian bootstrap for the error analysis of small-sample, high log-variance data sets compared to the standard bootstrap.

In addition to the continuous distributions just discussed, we also assessed the bootstrap performance on power law distributions, i.e. $x \sim x^{-\alpha}$, with $x > 0$ and $\alpha > 2$, and on normal distributions, i.e. $x \sim exp\frac{-(x-\mu)^2}{2\sigma^2}$, with $\sigma = 10$ and $\sigma = 1$. The results reveal that, despite a relatively small variance of sample means, the bootstrap can strongly under-estimate the true mean of a power law, mainly because of fairly small uncertainty ranges (Figures S5 and S6). On the other hand, data sampled from the normal distributions have mean values that are comparable to the true distribution mean, yielding error bars that cover it about as often as expected, even for very small data sets (Figures S7 and S8).

The ratio of the values of $\log(\bar{x}_{lo}/\mu)$ at half-max CDF (i.e. at CDF = 0.5) from the standard and the Bayesian bootstrap serves as a metric for the aforementioned opening gap or discrepancy in the corresponding lower limit CDFs of the two methods. Table 1 lists this ratio with the corresponding bootstrap performances (nominal 95% coverage) of small data sets ($n = 10$) along with other statistics derived numerically for all continuous distributions discussed in this study. The higher-order central moments measure the degree of deviation from the normal distribution in terms of asymmetry (skewness) or longer tails (excess kurtosis) and they usually assume a large value for distributions over many orders of magnitude (and are constant for the exponential distribution and zero for the normal distribution). Unlike the standard deviation of the distributions, $\sigma_x$, that of the logarithm of the values, $\sigma_{\log_{10}(x)}$, seems to be correlated with the corresponding half-max CDF ratio. In particular, when a strong bias of the standard bootstrap compared to the Bayesian bootstrap exists, i.e., when that ratio is > 1, we consistently find that $\sigma_{\log_{10}(x)}$ is > 1. This suggests that the value of $\sigma_{\log_{10}(x)}$ can be used as an indicator of the greater suitability of the Bayesian bootstrap over the standard method for the uncertainty estimation from small samples. Note that, $\sigma_{\log_{10}(x)}$ is defined and can be derived for power law distributions with $x > 0$ and $2 < \alpha < 3$, for which the calculation of higher moments does not converge.

As alternatives to the Bayesian bootstrap, we probed slight refinements to it that might improve its performance, i.e., reduce the degree of under-estimation. These are the Bayesian bootstrap with "pseudovalues", where an extra value is added to the synthetic data set before analysis, and the weighted Bayesian bootstrap, in which the sample mean value is weighted. However, the results were not promising, as described in the SI (see Figures S9 and S10).

In the future, the Bayesian bootstrap should also be compared to additional bootstrap methods, such as the bias-corrected percentile bootstrap, which aims to correct the bias of the standard bootstrap by manipulating the resampling procedure such that the bootstrap distributions are not centered as the empirical data set is.[27] Such a correction may be useful for providing reliable error bars for data sets, in which a small subset has a large impact on the distribution. Furthermore, the Bayesian bootstrap may be refined by specifying prior probability distributions that are based on some knowledge of the data.

**Rate constants from weighted ensemble simulations**

It is desirable to know how the standard and the Bayesian bootstrap perform on actual results from simulations or experiments. Computationally expensive simulations, such as the WE simulation of protein folding, may lead to a relatively small number of results that span multiple orders of magnitude. The folding rate constants in Table 2 range from 4.18e-17 to 7.35e+03 (System A) and from 3.87e-70 to 2.42e-01 (System B), respectively, exemplifying such small-size, high log-variance data sets that we are concerned with in this study. The given mean values, $\mu$, and the corresponding standard error of the mean, $se(\mu)$, indicate that the larger values in the two data sets dominate both statistics. We used these two data sets as the true distributions and we repeated the same comparative analysis as before (see Figure 1), i.e., we randomly generated synthetic data samples to evaluate the error analysis of the standard and the Bayesian bootstrap. This time, however, the synthetic data sets are of the same size as the original data sets, i.e., $n = 15$ (System A) and $n = 13$ (System B), and we assess the bootstrap performance at a nominal 95% coverage level.

Consistent with the results on the continuous distributions, the standard bootstrap confidence intervals for WE folding rate constants are biased toward very small values (Figure 5), whereas the Bayesian bootstrap is more robust toward this bias. Figure 5 shows the 95% confidence interval and credibility region of the mean values for synthetic data sets. As expected, there are instances in which the mean value, $\bar{x}$, is a few orders of magnitude smaller than the true mean. However, regardless of the value of $\bar{x}$, it is striking that the 95% standard bootstrap confidence intervals always span a wider segment than the 95% Bayesian credibility regions for both sets of protein folding rate constants.

Based on the subset of synthetic data sets (50 out of 1,000) shown in Figure 5, $\mu$ can be under- and over-estimated by both bootstrap methods. Recall that under-estimation indicates a confidence interval with an upper limit below the true mean and conversely for over-estimation and that both expected values for a 95% coverage is 2.5%. Averaged over all synthetic data sets ($N$ = 1,000), both bootstrap approaches under-estimate the true mean by 33% for System A and by 34% for System B. The Bayesian bootstrap also over-estimates the true mean ~7% of the time (for both systems) but the standard bootstrap only over-estimates in <1% of all cases.

As with the continuous distributions, the degree of bias toward very small values (increase of CDFs for $\log(\bar{x}_{lo}/\mu) < 0$) and the under-estimation of both methods (increase of CDFs up to ~33% for $\log(\bar{x}_{up}/\mu) < 0$) can be retrieved from the corresponding graphs shown in Figure 6. This error analysis of protein folding rate constants confirms that, unlike the standard bootstrap, the Bayesian bootstrap is not biased toward small data values, but its credibility region under-estimates the true mean just as often as the standard method does. Thus, the nominal 95% coverage range fails to cover the true mean in ~40% of all cases (under- plus over-estimations) or, in other words, it only corresponds to an effective ~60% coverage.

To set the data in context, Figure 7 shows how estimated rate constants for protein folding evolve over the course of a set of WE simulations. Theoretically, one expects rate estimates to start at $k_{fold} = 0$ sec$^{-1}$ at the beginning of a WE simulation and converge to a specific value over simulation time. Figure 7 shows such a graph for the average rate constant, which is dominated by larger rate constant values, along with the corresponding 95% confidence intervals from standard bootstrap and 95% credibility regions from the Bayesian bootstrap. Although, the upper limits of both uncertainty brackets are similar, the lower limits of the credibility region are more appropriate than those of the confidence interval, as shown by the analysis in this study.

**Conclusions**

The proper assessment of uncertainty is fundamental in data analysis in order to reach reliable conclusions. In the field of molecular simulations, results that are highly variable over orders of magnitude and estimated with small samples are an intrinsic consequence of certain types of calculations - and these require a different method of uncertainty estimation than the conventional standard error of the mean. Although the standard and Bayesian bootstrap are similar as they both assess the uncertainty of the mean value through a resampling procedure, they differ fundamentally in following a frequentist or a Bayesian approach. The standard bootstrap characterizes expected variation in *an observable* based on a fixed approximate distribution, whereas the Bayesian bootstrap estimates variation in the *underlying distribution* of assumed multinomial form. Conceptually, the primary goal would seem to be the accurate characterization of the underlying distribution which governs all observables.

The comparison of the two methods on random data sets sampled from different common continuous distributions and on data sets of rate constants from weighted ensemble simulations shows that the Bayesian bootstrap yields a significantly more reliable credibility region for the mean value. That is, the Bayesian approach covers the true mean about as often as the standard approach but with *a much smaller range* in the log space most pertinent for rate estimation. This is a significant result because it suggests that the fundamental question to ask for assessing uncertainty should be 'Based on the single set of observed data, what degree of variation is likely for the underlying distribution?' rather than 'How much variation do we expect in observed data based on assuming a single underlying distribution?' It must be noted, however, that the uncertainty ranges produced by both bootstrap methods can significantly under-estimate the true mean of an underlying data set. This under-estimation of the true mean, which is a consequence of fairly small sample sizes, has been reported before for the standard percentile bootstrap method.[26-27] The systematic analysis of the degree of bootstrap under- and over-estimation at different uncertainty levels and with increasing data sets has shed light on this failure to cover the true mean of a population as a result of small sample sizes from high log-variance distributions.

The usefulness of the continuous distributions for this study can be understood when comparing the data in Figure 2A to those of the simulation rate constants in Figure 5A, which are most similar among all studied distributions. Both are spread over 20 orders of magnitude and have comparable synthetic data set sizes ($n = 10$ and $n = 13$). However, due to the relatively small amount of available rate constants from simulation, the accuracy with which the uncertainty of the bootstrap methods could be assessed was limited. In fact, based on the 1,000-fold synthetic data sampling, a nominal 95% coverage led to an effective ~60% coverage when applied to the rate constant data but only an effective ~42% coverage was reached in the case of the continuously distributed data. This reveals the value of using continuous distributions for the rigorous analysis of the bootstrap performance. Moreover, the detailed comparison of the two bootstrap methods applied to continuous distributions with different degrees of log variance revealed that the standard deviation of the logarithm of the distribution data serves as a reliable indicator for when the Bayesian bootstrap performs better.

This study reinforces the lesson that error analysis should be performed with caution.[23, 31] Certain data sets may not be amenable to conventional techniques because of their spread and small quantity of sampled data, which often results from computationally expensive calculations.

Bootstrapping is a reasonable approach, that does not make any prior assumptions,[25] for the uncertainty estimation of such small-sample, high log-variance distributions. In particular, as we have shown in this study, the Bayesian bootstrap is even better suited for this problem than the standard bootstrap.

**Acknowledgements**

We thank Upendra Adhikari and Jeremy Copperman for helpful discussions and we gratefully acknowledge support from the NIH (Grant GM115805) and from the OHSU Center for Spatial Systems Biomedicine.

**Figures**

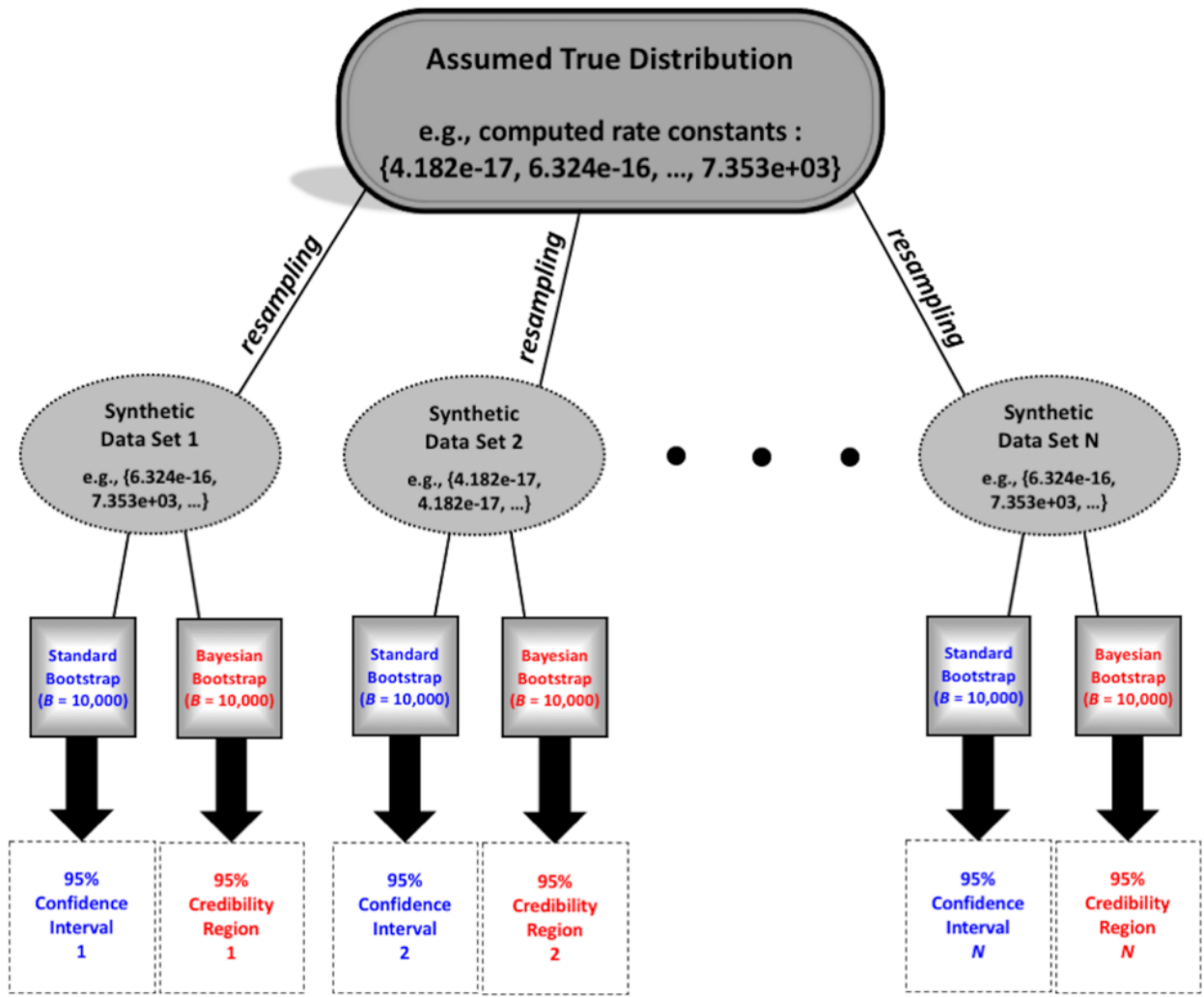

Figure 1

Synthetic data generation and error analysis. The assumed underlying true distribution can be any distribution of interest, e.g., the folding rate constants from WE simulations (Table 2). $N$ synthetic data sets (with $N = 1,000$) are generated by resampling, i.e. drawing with replacement, from the true distribution and the two bootstrap procedures are applied to each data set with $B = 10,000$ replications in both. For each synthetic data set a 95% confidence interval and a 95% credibility region is obtained by the standard and by the Bayesian bootstrap, respectively.

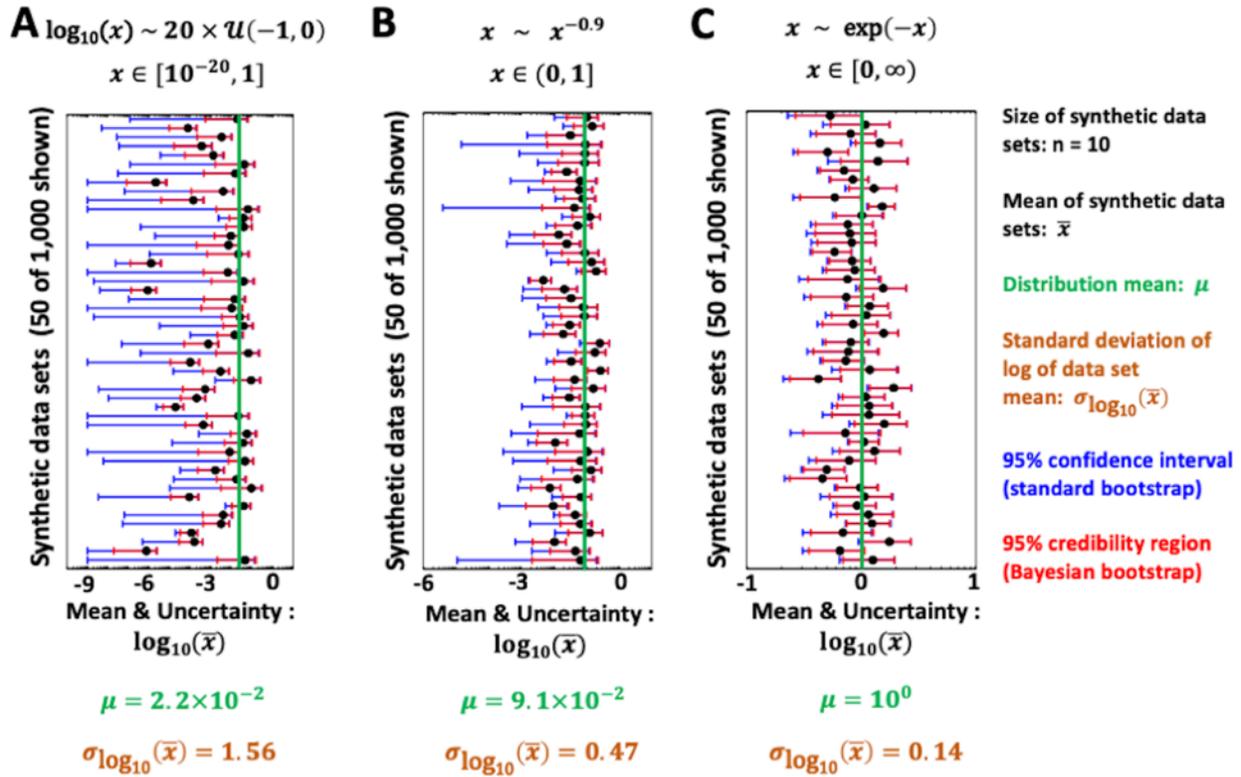

Figure 2

Comparison of standard and Bayesian bootstrap uncertainty ranges for data samples from three continuous distributions (A, B, and C). For 50 out of 1000 synthetic data sets, the standard bootstrap 95% confidence interval (blue) and the 95% Bayesian credibility region (red) are shown, which should be compared with the true mean (vertical green line). For reference the sample mean (black circle) of each synthetic data set, $\bar{x}$, is also shown. The standard deviation of the logarithm of the sample means, $\sigma_{\log_{10}(\bar{x})}$, is a measure of the log variance of the underlying ("true") distribution.

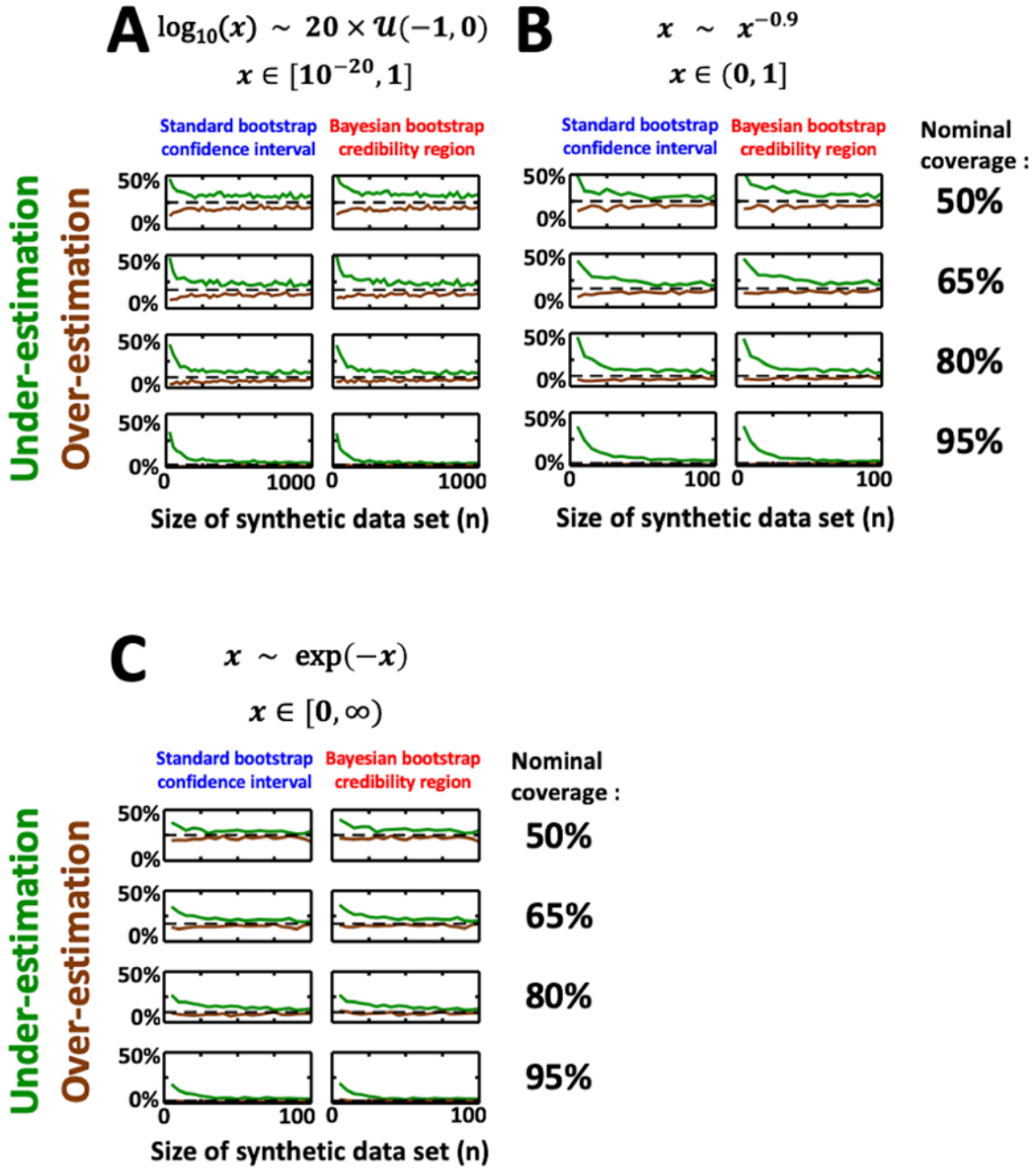

Figure 3

Failure to cover true mean: under and over-estimation. The under- and over-estimation percentage of the standard bootstrap confidence interval (left column) and the Bayesian bootstrap credibility region (right column) as a function of data set size shown for the three continuous distributions (A, B, and C) at different nominal coverage ranges. The expected under- and over-estimations, which are 100% minus half of the nominal coverage, are shown as dashed lines.

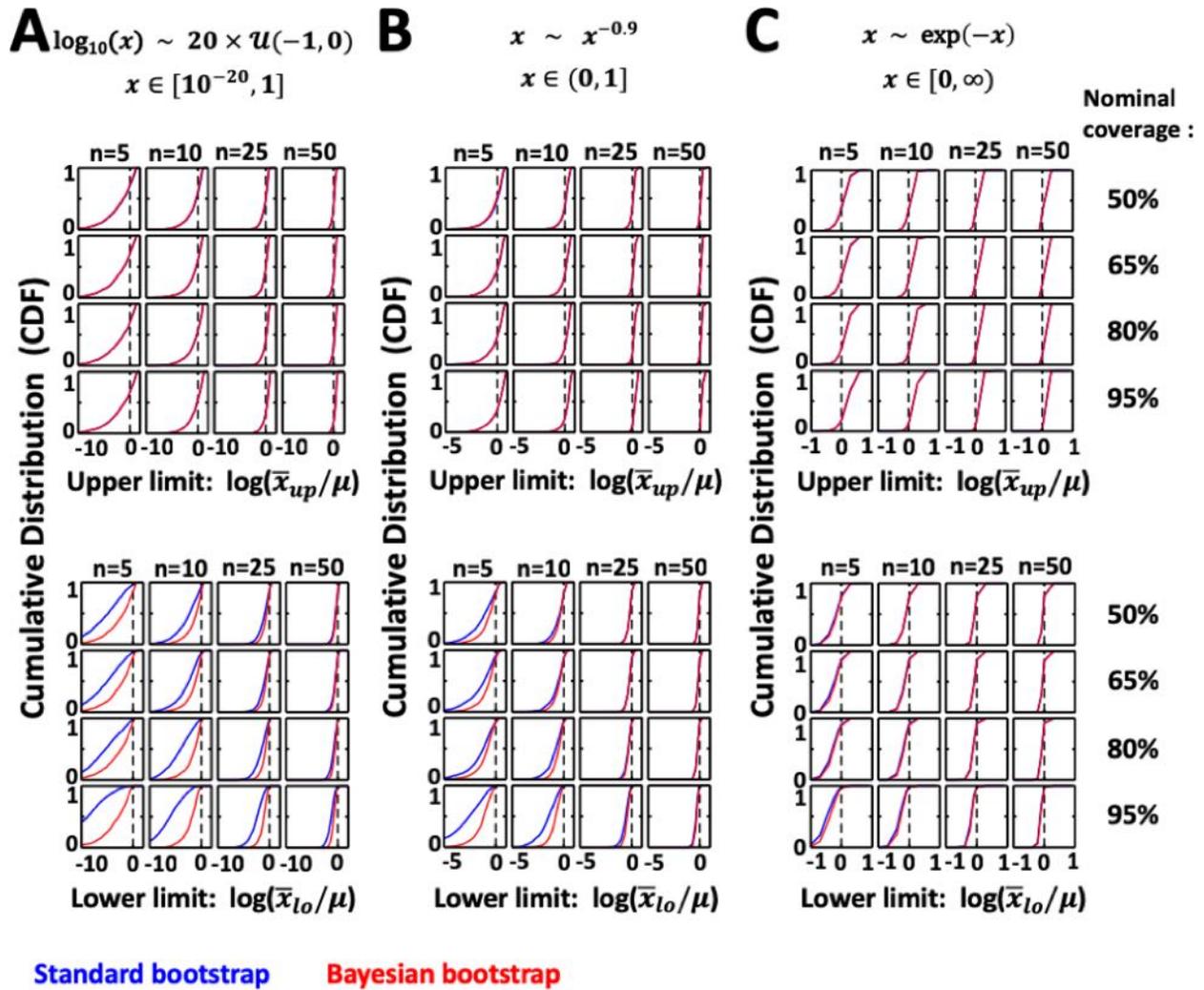

Figure 4

Comparison of upper and lower limits of standard (blue) and Bayesian ranges (red) for four relatively small data samples ($n$ = 5, 10, 25, 50) shown for the three continuous distributions (A, B, and C) at different nominal coverage ranges. The plots show cumulative distributions (CDFs) for upper limits ($\bar{x}_{up}$) and lower limits ($\bar{x}_{lo}$) using their ratio to the true mean ($\mu$). Vertical dashed lines are reference values of when the upper or lower limits correspond to the true mean, i.e., when under- or over-estimations can occur (note the log scale on the horizontal axis).

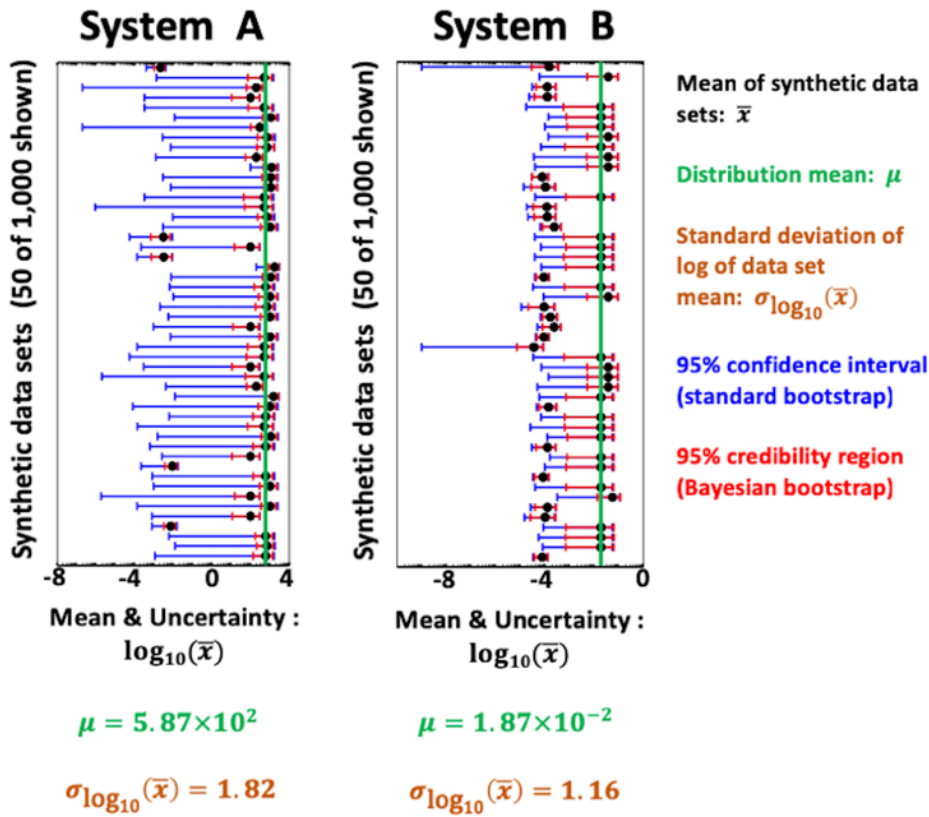

Figure 5

Comparison of standard and Bayesian bootstrap uncertainty ranges. For 50 out of 1000 synthetic data sets, the standard bootstrap 95% confidence interval (blue) and 95% Bayesian credibility region (red) are shown, which should be compared with the true mean (vertical green line). Two systems (A and B) are shown – see Table 2. For reference the sample mean (black circle) of each synthetic data set, $\bar{x}$, is also shown. The standard deviation of the logarithm of the sample means, $\sigma_{\log_{10}(\bar{x})}$, is a measure of the log variance of the underlying ("true") distribution.

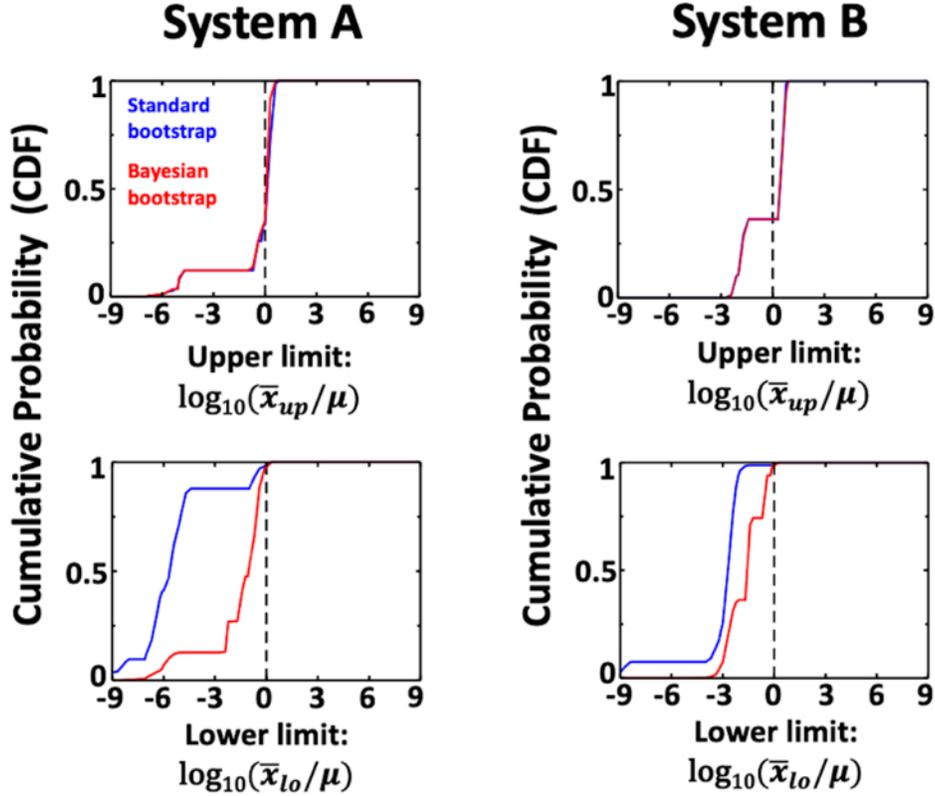

Figure 6

Comparison of upper and lower limits of standard and Bayesian ranges. The plots show cumulative distributions (CDFs) for upper limits ($\bar{x}_{up}$) and lower limits ($\bar{x}_{lo}$) of the standard bootstrap 95% confidence interval (blue) and Bayesian 95% credibility region (red) using their ratio to the true mean ($\mu$) for the two systems (A and B). Vertical dashed lines are reference values of when the upper or lower limits correspond to the true mean, i.e., when under- or over-estimations can occur (note the log scale on the horizontal axis). The two approaches perform very similarly for the upper limits but the standard bootstrap most often yields lower limits which are multiple orders of magnitude smaller than the true mean.

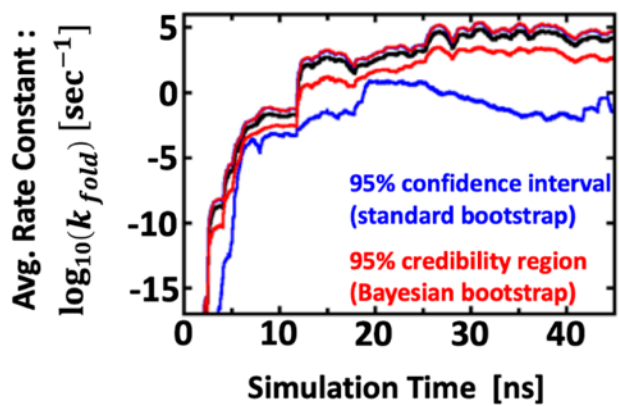

Figure 7

Example data from molecular simulation. Estimates for a protein folding rate constant, $k_{fold}$, (black) versus the simulation time averaged over 30 individual weighted ensemble simulations along with the upper and lower limits of the 95% confidence interval (blue) from the standard bootstrap and the 95% credibility region (red) from the Bayesian bootstrap (adapted from Adhikari et al.[16]).

Table 1

Distribution metrics and bootstrap performance for all continuous distributions discussed in this study. The standard deviation, $\sigma_x$, that of the logarithm of values, $\sigma_{\log_{10}(x)}$, the skewness, $\mu_{(3)}/\sigma_x^3$, and the excess kurtosis, $\mu_{(4)}/\sigma_x^4 - 3$, where $\mu_{(n)} = \int dx\, p(x)(x-<x>)^n$ is the $n^{\text{th}}$ central moment, are shown for each distribution together with the two bootstrap performances (actual coverage of a nominal 95% coverage) and the half-max CDF ratio of the two bootstrap methods applied to a small data set of size $n = 10$.

| Distribution Type | Parameter | $\sigma_x$ | $\sigma_{\log(x)}$ | Skewness | Excess kurtosis | Standard bootstrap coverage | Bayesian bootstrap coverage | Half-max CDF Ratio |
|---|---|---|---|---|---|---|---|---|
| $\log_{10}(x) \sim k \times [-1, 0]$ | $k = 20$ | 0.10 | 5.78 | 6.2 | 45.0 | 44.2% | 44.3% | $2 \times 10^4$ |
| | $k = 5$ | 0.19 | 1.44 | 2.8 | 10.4 | 74.1% | 74.5% | 2.5 |
| $x \sim x^{-\alpha}$ $x \in (0, 1]$ | $\alpha = 0.9$ | 0.20 | 4.34 | 2.7 | 9.6 | 75.7% | 76.1% | 5.0 |
| | $\alpha = 0.1$ | 0.29 | 0.48 | 0.1 | 1.8 | 92.3% | 91.8% | 1.0 |
| $x \sim x^{-\alpha}$ $x \in (0, \infty)$ | $\alpha = 2.9$ | not defined | 0.23 | not defined | not defined | 72.1% | 71.4% | 1.0 |
| | $\alpha = 2.1$ | not defined | 0.40 | not defined | not defined | 26.3% | 27.0% | 1.0 |
| $x \sim \exp(-\lambda x)$ | $\lambda = 10^0$ | 1 | 0.56 | 2.0 | 6.0 | 85.5% | 84.5% | 1.0 |
| | $\lambda = 10^{-6}$ | $10^6$ | 0.56 | 2.0 | 6.0 | 86.4% | 85.7% | 1.0 |
| $x \sim \exp\dfrac{-(x-\mu)^2}{2\sigma^2}$ ($\mu = 30$) | $\sigma = 10$ | 10 | 0.19 | 0.0 | 0.0 | 89.9% | 88.3% | 1.0 |
| | $\sigma = 1$ | 1 | 0.01 | 0.0 | 0.0 | 89.6% | 89.1% | 1.0 |

Table 2

Folding rate constants for two different proteins (Systems A and B), obtained from 15 (A) and 13 (B) independent WE simulations, are spread over 20 (A) and 69 (B) orders of magnitude, respectively. Both the mean value, µ, and the standard error of the mean, se(µ), are about the same magnitude as the largest data value.

| Simulation | System A | System B |
|---|---|---|
| 1 | 4.182e-17 | 3.870e-70 |
| 2 | 6.324e-16 | 9.975e-27 |
| 3 | 1.126e-14 | 4.649e-22 |
| 4 | 1.907e-12 | 2.423e-20 |
| 5 | 1.797e-10 | 2.470e-19 |
| 6 | 8.206e-08 | 3.808e-11 |
| 7 | 1.500e-07 | 1.147e-10 |
| 8 | 1.610e-06 | 4.739e-05 |
| 9 | 1.303e-05 | 1.299e-04 |
| 10 | 8.027e-04 | 1.706e-04 |
| 11 | 1.203e-03 | 2.579e-04 |
| 12 | 9.201e-03 | 9.113e-04 |
| 13 | 4.283e-02 | 2.415e-01 |
| 14 | 1.449e+03 | – |
| 15 | 7.353e+03 | – |
| µ | 5.868e+02 | 1.870e-02 |
| se(µ) | 4.928e+02 | 1.857e-02 |

# Statistical uncertainty analysis for small-sample, high log-variance data: Cautions for bootstrapping and Bayesian bootstrapping

## –

## Supporting Information

Barmak Mostofian and Daniel M. Zuckerman[*]

Department of Biomedical Engineering, School of Medicine,
Oregon Health & Science University, Portland, Oregon

[*]zuckermd@ohsu.edu

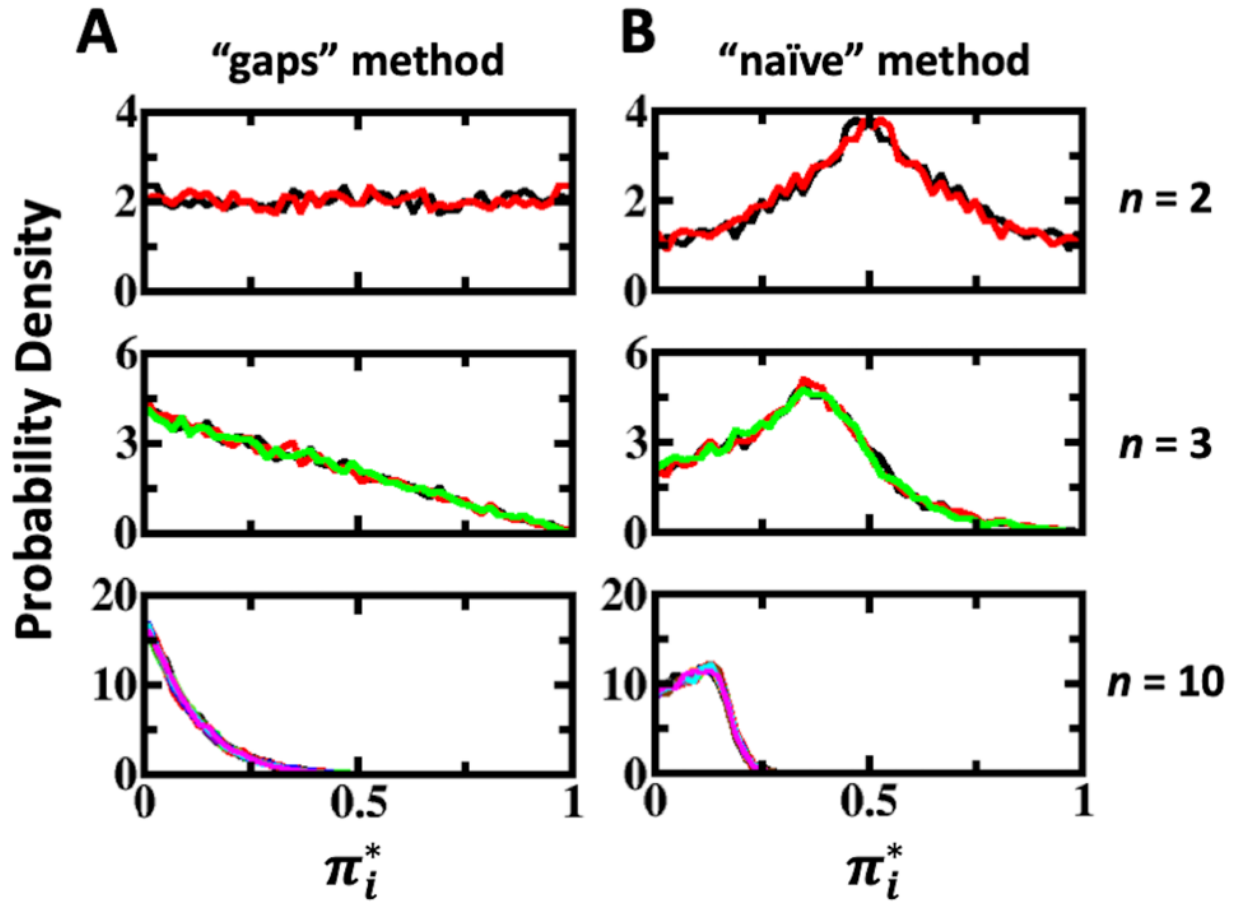

Figure S1

The distribution of $n$ different values of $\pi_i^*$ with $\sum_{i=1}^{n} \pi_i^* = 1$, for $n = 2, 3,$ and 10, generated (A) by the "gaps" method, in which the $n$ values are derived as the differences (or gaps) between succeeding pairs in an ordered list of $n+1$ random real numbers between 0 and 1, and (B) by the "naïve" method of simply normalizing $n$ uniform random variates. The $n$ different colors correspond to the $n$ different distributions. Each probability density distribution is based on 10,000 repeats.

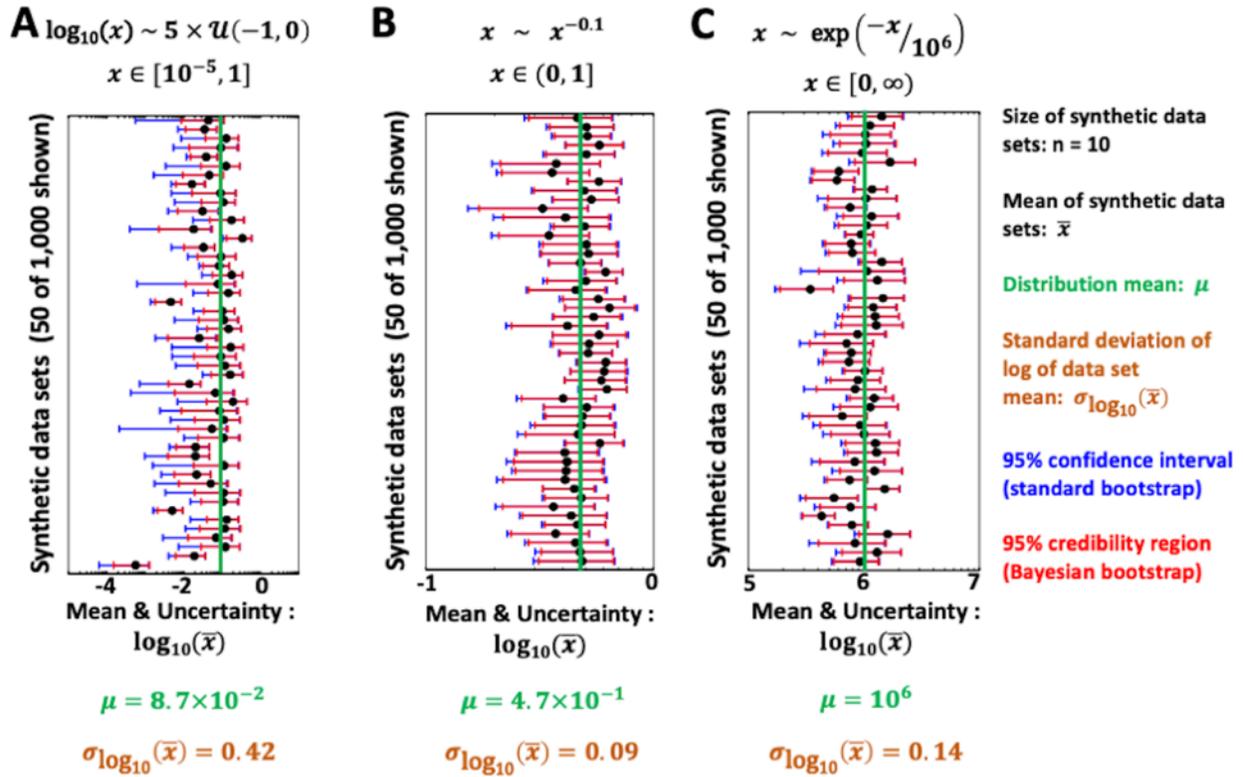

Figure S2

Comparison of standard and Bayesian bootstrap uncertainty ranges for data samples from three continuous distributions with smaller values for the parameters than those presented in Figure 2, i.e. with $k = 5$ (A), $\alpha = 0.1$ (B), and $\lambda = 10^6$ (C). For 50 out of 1000 synthetic data sets, the standard bootstrap 95% confidence interval (blue) and the 95% Bayesian credibility region (red) are shown, which should be compared with the true mean (vertical green line). For reference the sample mean (black circle) of each synthetic data set, $\bar{x}$, is also shown. The standard deviation of the logarithm of the sample means, $\sigma_{\log_{10}(\bar{x})}$, is a measure of the log variance of the underlying ("true") distribution.

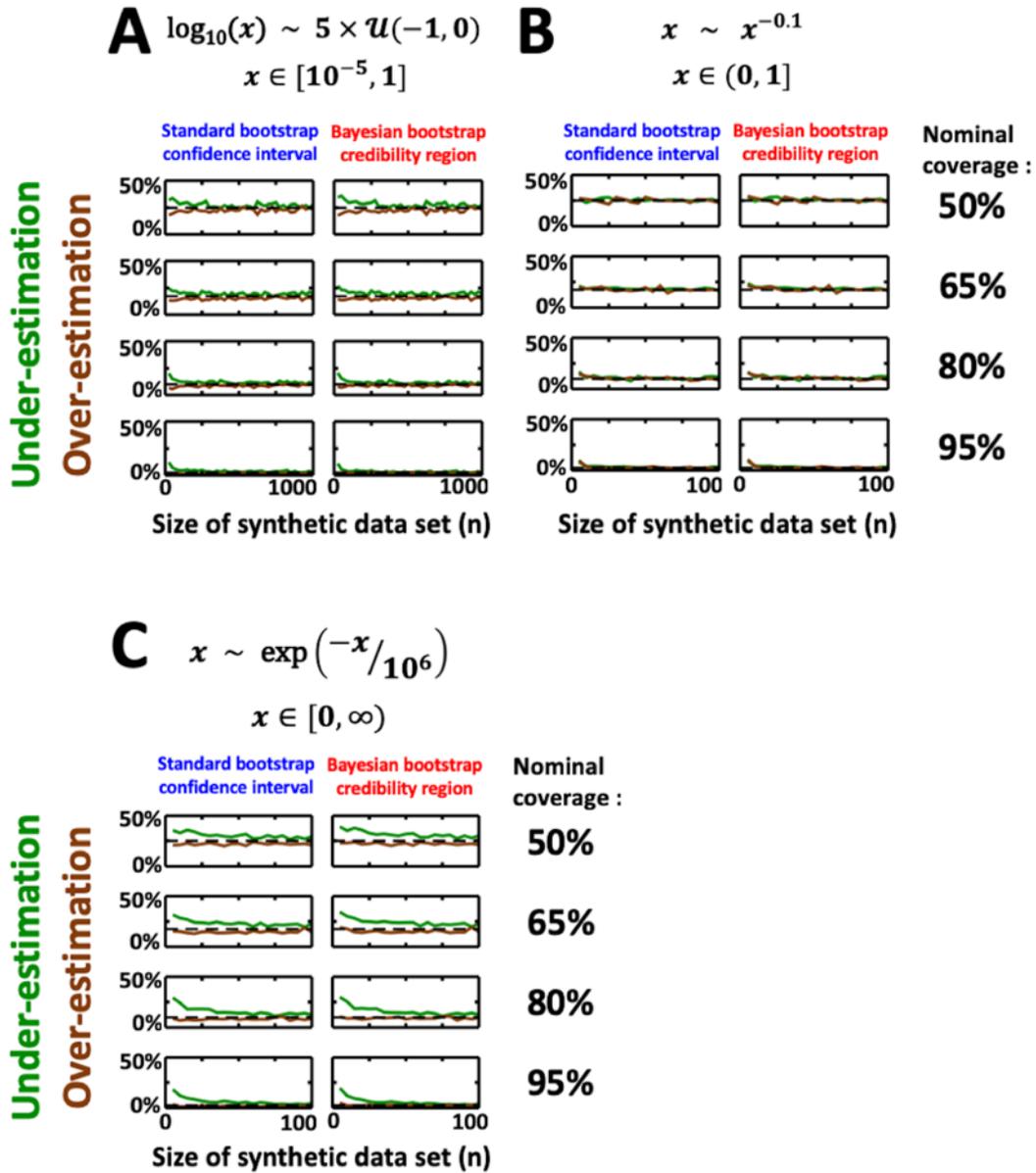

Figure S3

The under- and over-estimation percentage of the standard bootstrap confidence interval (left column) and the Bayesian bootstrap credibility region (right column) as a function of data set size at different nominal coverage ranges shown for the three continuous distributions with smaller values for the parameters than those presented in Figure 3, i.e. with $k = 5$ (A), $\alpha = 0.1$ (B), and $\lambda = 10^6$ (C). The expected under- and over-estimations, which are 100% minus half of the nominal coverage, are shown as dashed lines

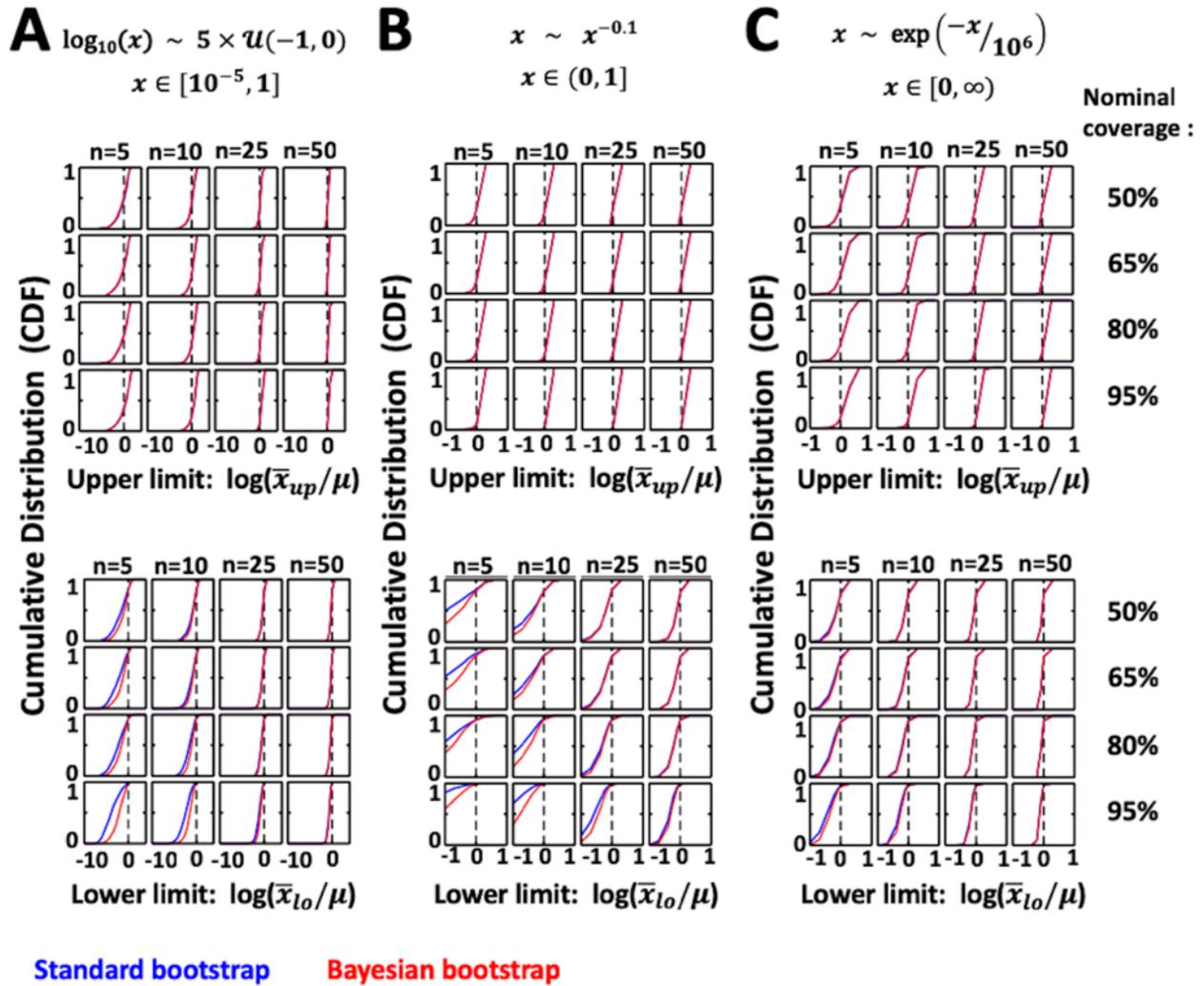

Figure S4

Comparison of upper and lower limits of standard (blue) and Bayesian ranges (red) for four relatively small data samples at different nominal coverage ranges shown for the three continuous distributions with smaller values for the parameters than those presented in Figure 4, i.e. with $k = 5$ (A), $\alpha = 0.1$ (B), and $\lambda = 10^6$ (C). The plots show cumulative distributions (CDFs) for upper limits ($\bar{x}_{up}$) and lower limits ($\bar{x}_{lo}$) using their ratio to the true mean ($\mu$).

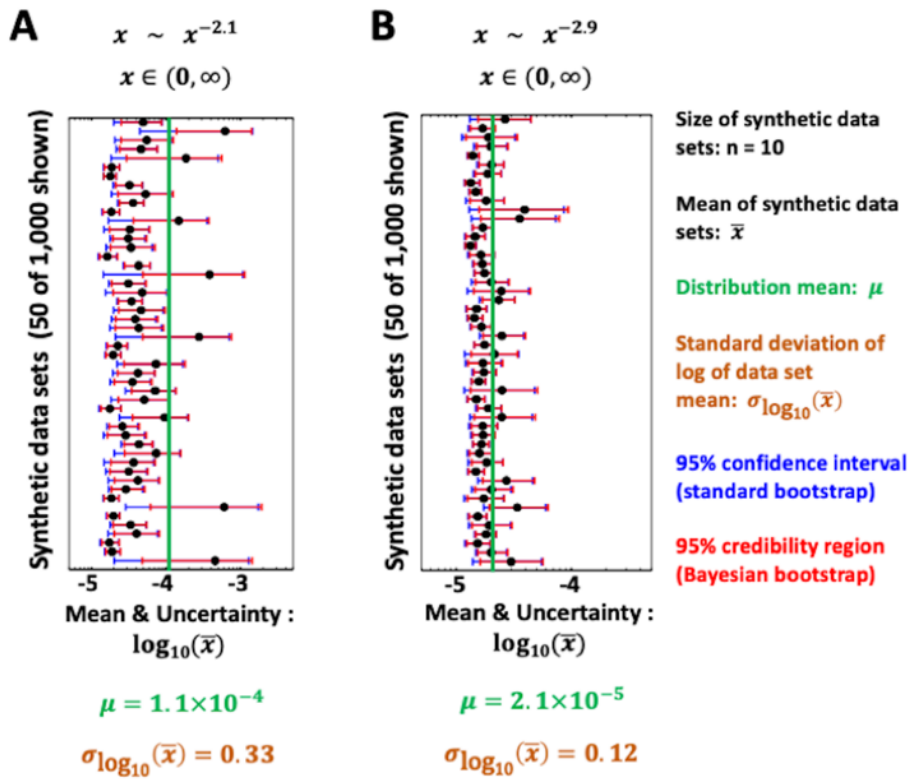

Figure S5

Comparison of standard and Bayesian bootstrap uncertainty ranges for data samples from power law distributions with $x > 0$ and $\alpha = 2.1$ (A) or $\alpha = 2.9$ (B). For each synthetic data set, the standard bootstrap 95% confidence interval (blue) and the 95% Bayesian credibility region (red) are shown, which should be compared with the true mean (vertical green line).

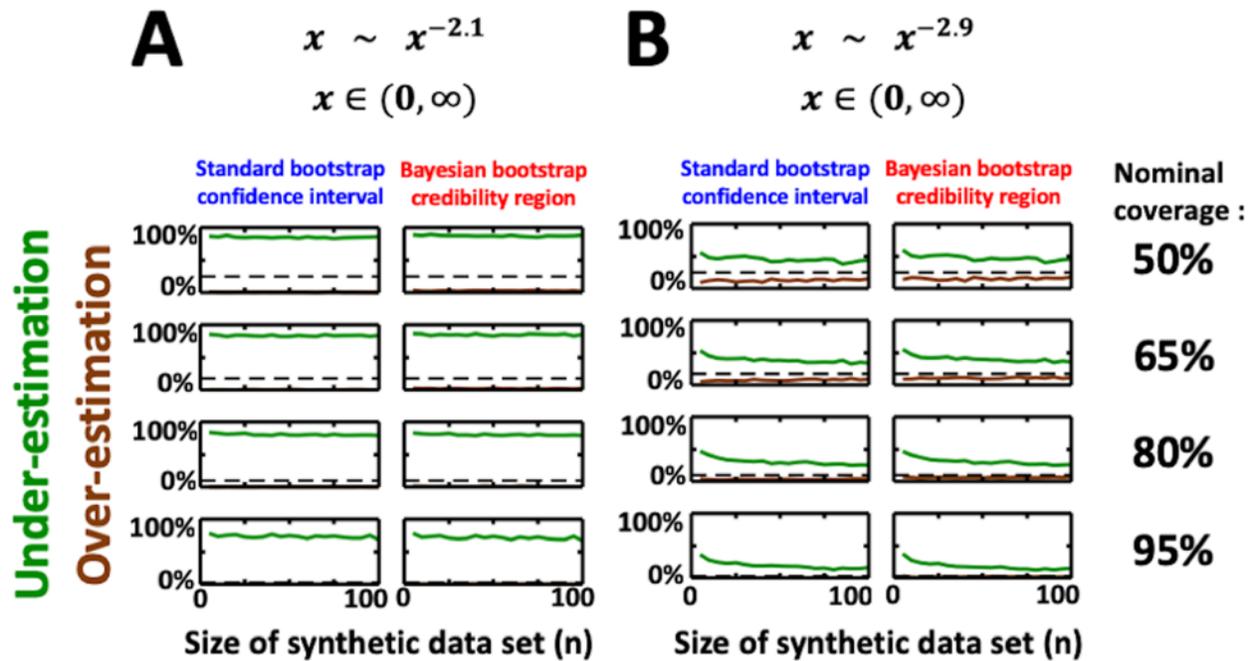

Figure S6

The under- and over-estimation percentage of the standard bootstrap confidence interval (left column) and the Bayesian bootstrap credibility region (right column) as a function of data set size shown for the power law distributions with $x > 0$ and $\alpha = 2.1$ (A) or $\alpha = 2.9$ (B). The expected under- and over-estimations, which are 100% minus half of the nominal coverage, are shown as dashed lines

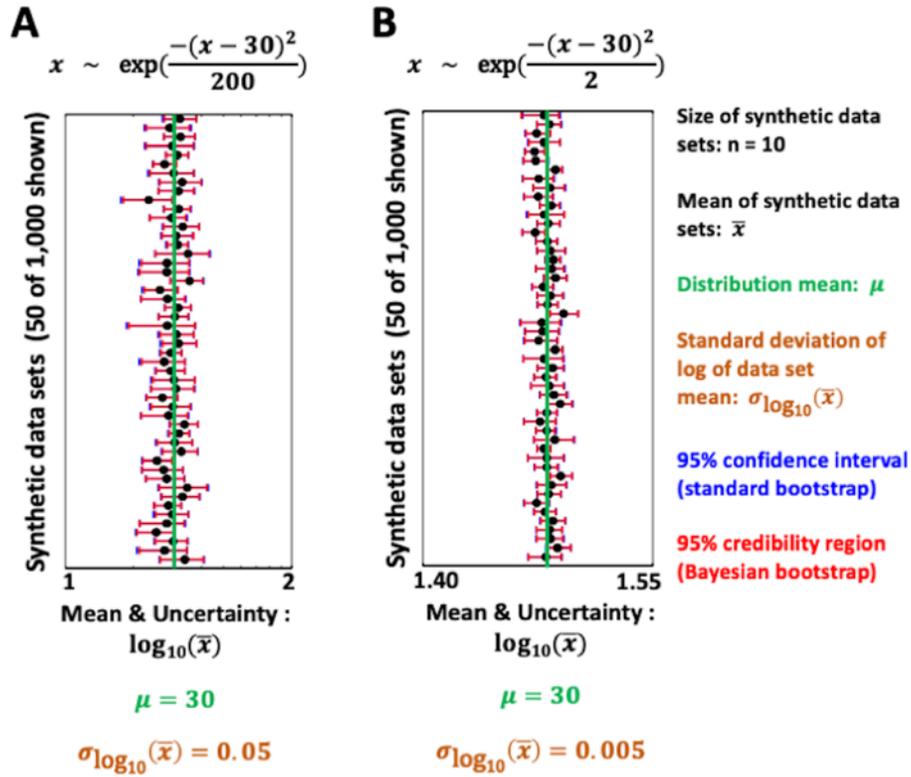

Figure S7

Comparison of standard and Bayesian bootstrap uncertainty ranges for data samples from normal distributions with $\sigma = 10$ (A) or $\sigma = 1$ (B). For each synthetic data set, the standard bootstrap 95% confidence interval (blue) and the 95% Bayesian credibility region (red) are shown, which should be compared with the true mean (vertical green line). Note that the mean of the normal distribution has been set to $\mu = 30$ so that no negative values are sampled under the given values for $\sigma$.

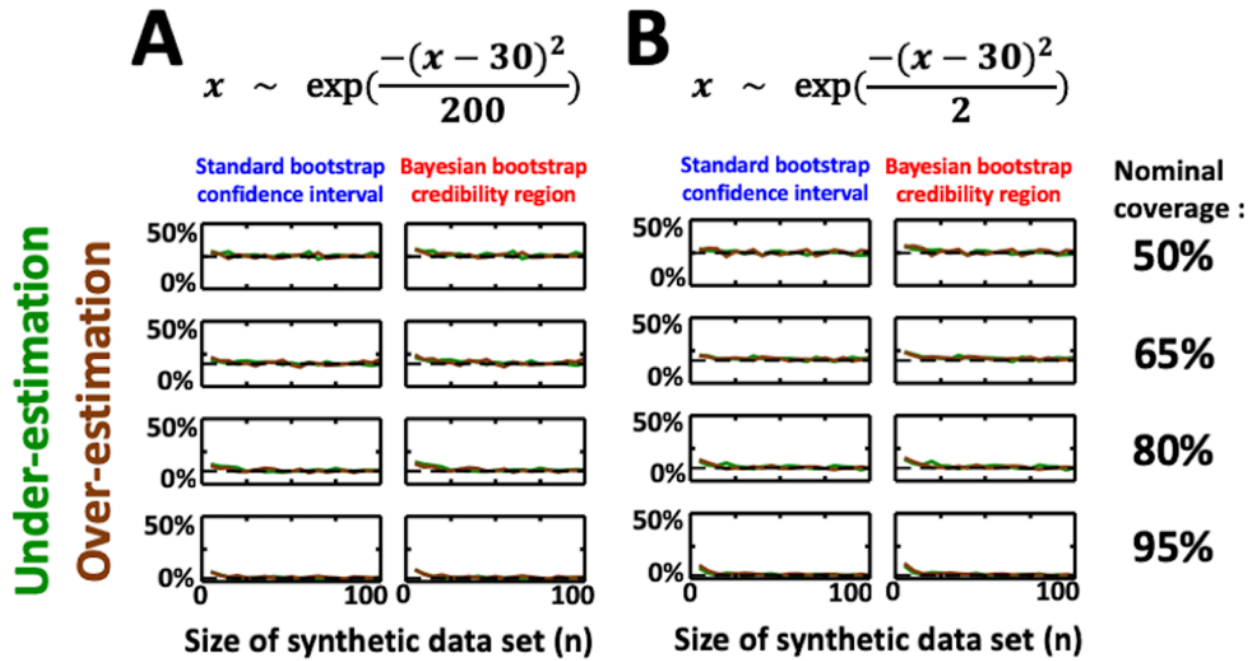

Figure S8

The under- and over-estimation percentage of the standard bootstrap confidence interval (left column) and the Bayesian bootstrap credibility region (right column) as a function of data set size shown for the normal distributions with $\sigma = 10$ (A) or $\sigma = 1$ (B). The expected under- and over-estimations, which are 100% minus half of the nominal coverage, are shown as dashed lines

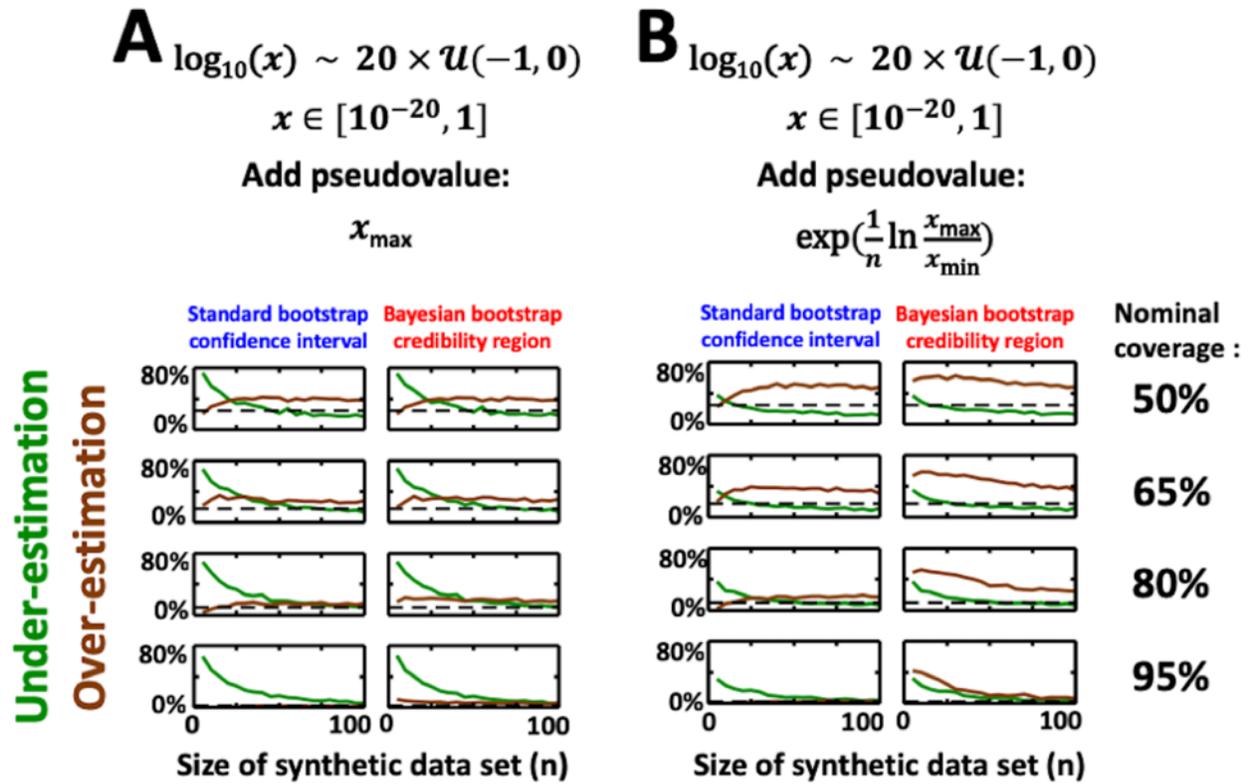

Figure S9

The under- and over-estimation percentage of the standard bootstrap confidence interval (left column) and the Bayesian bootstrap credibility region (right column) as a function of data set size shown for the continuous distribution of scalar data with logarithms within $[-20, 0]$ at different nominal coverage ranges and for data sets to which an extra value (or pseudovalue) was added in order to possibly reduce the percentage of under-estimations observed in Figure 5A. The pseudovalue is either the maximum value of the sample, $x_{max}$ (A), or the scaled maximum value of the sample, $\exp(\frac{1}{n}\ln\frac{x_{max}}{x_{min}})$ (B), which corresponds to a value with a normalized exponent between that of $x_{min}$ and $x_{max}$. The expected under- and over-estimations are shown as dashed lines. The addition of either pseudovalue to the data samples resulted in a strong over-estimation of the true mean (40-50%), particularly for small data sets of the Bayesian bootstrap, and a drop of the under-estimation below its expected percentage at lower nominal coverages (50-80%).

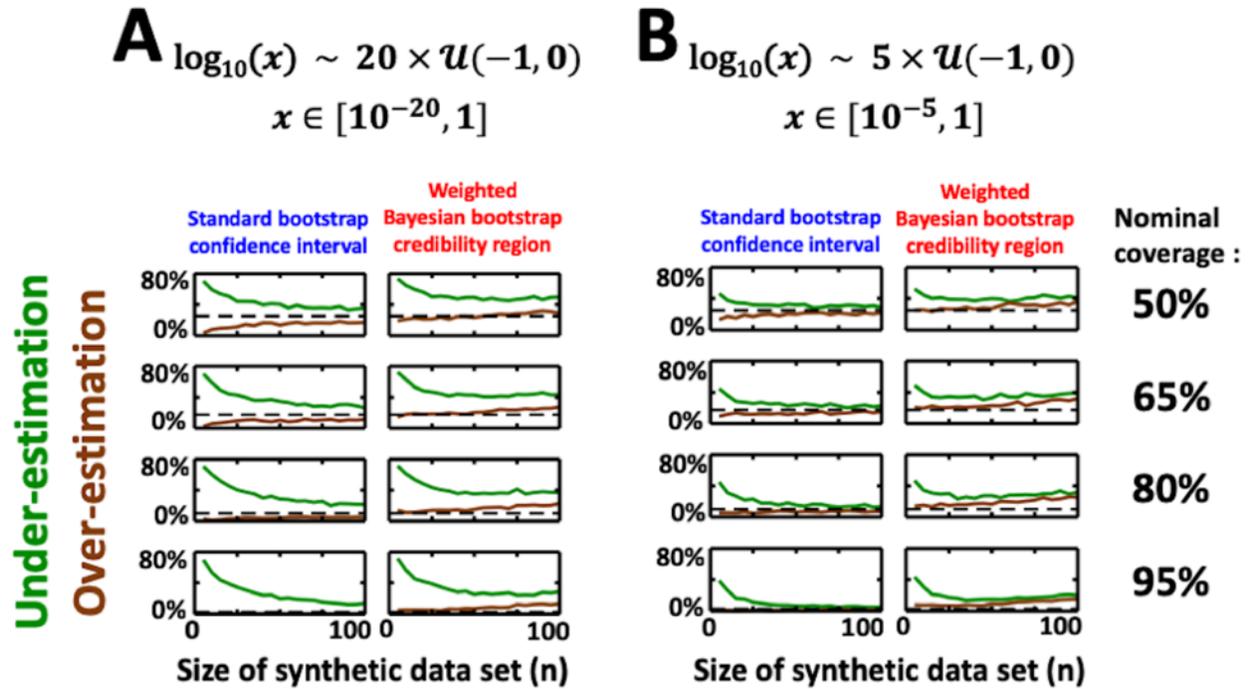

Figure S10

The under- and over-estimation percentage of the standard bootstrap confidence interval (left column) and a weighted Bayesian bootstrap credibility region (right column) as a function of data set size shown for the continuous distribution of scalar data with logarithms within $[-20, 0]$ (A) and within $[-5, 0]$ (B) at different nominal coverage. The expected under- and over-estimations are shown as dashed lines. In the weighted Bayesian bootstrap, each sample mean value $\bar{x}^*$ is weighted by a sample posterior likelihood $\mathcal{L}^* = \prod_{i=1}^{n} \pi_i^*$, which effectively corresponds to using a flat (uninformative) prior for the parameters $\pi^*$. (Operationally, this means that the credibility region is obtained from a weighted cumulative distribution function, $CDF_{\bar{x}^*}$.) We found that this causes both the under- and over-estimation to be constantly larger (by up to 30%) than the corresponding expected values.